\documentclass[printer]{aa}

\usepackage{graphicx}
\usepackage{natbib}
\usepackage{footnote}
\bibpunct{(}{)}{;}{a}{}{,}

\usepackage{txfonts}
\begin{document}

   \title{Kinematic study of the molecular gas associated with two cometary globules in Sh2$-$236}


   \author{M. E. Ortega
          \inst{1}
          \and
          S. Paron \inst{1,}\inst{2}
          \and
          M. B. Areal \inst{1}
          \and
          M. Rubio \inst{3}
          }

\institute{CONICET - Universidad de Buenos Aires, Instituto de Astronom\'{\i}a y F\'{\i}sica del Espacio (IAFE),
             CP 1428 Buenos Aires, Argentina\\
             \email{mortega@iafe.uba.ar}
\and Universidad de Buenos Aires, Facultad de Arquitectura, Dise\~{n}o y Urbanismo, Buenos Aires, Argentina
\and Departamento de Astronom\'{\i}a, Universidad de Chile, Casilla 36-D, Santiago, Chile
}

   \date{Received ; accepted }

 
  \abstract
  {}
   {Cometary globules, dense molecular gas structures exposed to the UV radiation, are found inside \ion{H}{ii}~regions. Understanding the nature and origin of these structures through a kinematic study of the molecular gas could be useful to advance in our knowledge of the interplay between radiation and molecular gas.}
   { Using the Atacama Submillimeter Telescope Experiment (ASTE; Chile) we carried out molecular observations towards two cometary globules in the \ion{H}{ii}~region Sh2$-$236. We mapped two regions of about $1\arcmin \times 1\arcmin$ with the $^{12}$CO J=3$-$2 and HCO$^+$ J=4$-$3 lines.  Additionally,  we carried out two single pointings with the C$_2$H, HNC, and HCN J=4$-$3 transitions. The angular resolution was about 22$\arcsec$. We combined our molecular observations with public infrared and optical data to analyse the distribution and kinematics of the molecular gas.}
  {We found kinematic signatures of infalling gas in the $^{12}$CO J=3$-$2 and C$_2$H J=4$-$3 spectra towards Sim 129 . 
We detected HCO$^+$, HCN, and HNC J=4$-$3 only towards Sim 130. The HCN/HNC integrated ratio of about 3 found in Sim~130 suggests that the possible star formation activity inside this globule has not yet ionized the gas. The location of the NVSS source 052255$+$33315, which peaks towards the brightest border of the globule, supports this scenario. The non-detection of these molecules towards Sim 129 could be due to the radiation field arising from the star formation activity inside this globule.  The ubiquitous presence of the C$_2$H molecule towards Sim 129 and Sim 130 evidences the action of the nearby O-B stars irradiating the external layer of both globules. 
Based on the mid-infrared 5.8~$\mu$m emission, we identified two new structures: (1) a region of diffuse emission (R1) located, in projection, in front of the head of Sim 129  and (2) a pillar-like feature (P1) placed besides Sim 130. 
Based on the $^{12}$CO J=3$-$2 transition, we found molecular gas associated with Sim 129, Sim 130, R1 and P1 at radial velocities of $-$1.5~kms$^{-1}$, $-$11~kms$^{-1}$, $+$10~kms$^{-1}$, and $+$4~kms$^{-1}$, respectively. Therefore, while Sim 129 and P1 are located at the far side of the shell, Sim 130 is placed at the near side consistent with earlier results.
Finally, the molecular gas related to R1 exhibits a radial velocity that differs in more than 11~km~s$^{-1}$ with the radial velocity of S129, which suggests that while S129 is located at the far side of the expanding shell, R1 would be placed  well beyond.}
   {}


   \maketitle
%

\section{Introduction}

It is well known that massive O-B stars originate the \ion{H}{ii}~regions, which during their expansion compress and shape the material that surrounds them. Different kind of structures have been observed at the inner surface of the resulting dusty molecular shells. For instance, dense clumps at the interface between the \ion{H}{ii} region and the cloud, pillars of gas pointing towards the ionizing sources, and globules detached from the parental molecular cloud. The cometary globules (CGs) are among these last features. They are small and dense clouds consisting of a dense head, surrounded by a bright rim, and prolonged by a diffuse tail. 

Several mechanisms have been suggested to explain the formation of this kind of structures, such as collect and collapse \citep{elme77}, radiation-driven implosion \citep[e.g.][]{san82, ber89, lef94, bis11}, shadowing \citep[][]{cerq06, mac10}, and collapse due to the shell curvature \citep{trem12b}. 
However, this is a matter still under debate. Understanding the origin of these structures would be useful to advance in our knowledge of the  interplay between radiation and molecular gas. 

The initial morphology and the clumpiness of the molecular cloud play a key role in the formation of these structures, since they can generate deformations and curvatures in the shell during the expansion of the \ion{H}{ii} region \citep{wal13}. Numerical simulations have shown how pillars can arise from the collapse of the shell \citep[e.g,][]{grits10, trem12b}, and how globules can be formed from the interplay between a turbulent molecular cloud and the ionization from massive stars \citep{trem12}.

\begin{figure}
\centering
 \includegraphics[width=9.0cm]{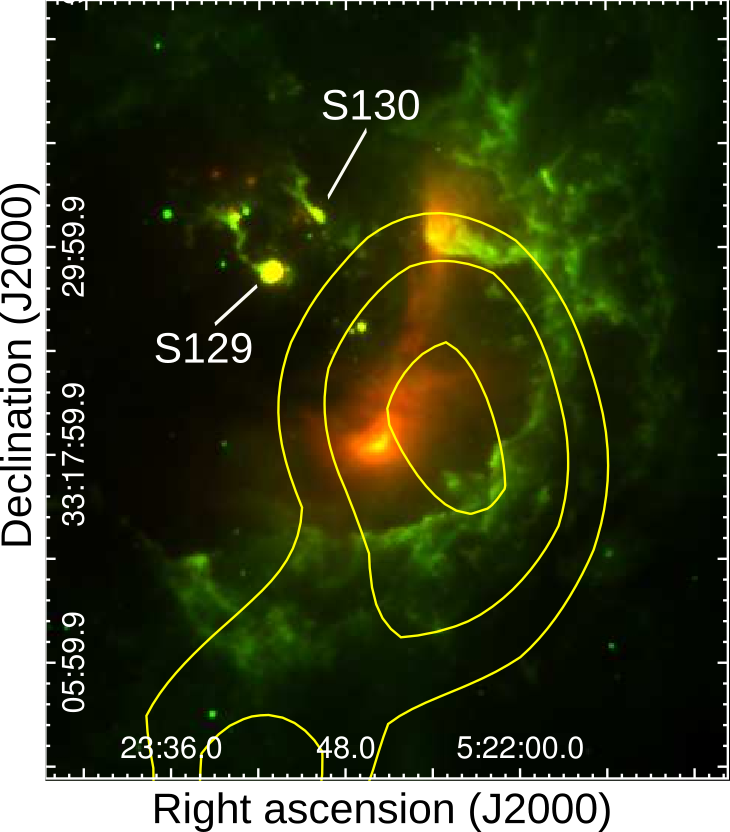}
 \caption{WISE two-color image of the \ion{H}{ii} region Sh2-236 and the cometary globules S129 and S130 at 12~$\mu$m in green and 22~$\mu$m in red. The yellow contours represent the $^{12}$CO (1-0) emission (angular resolution about 3.75 arcmin) integrated between $-10$ km s$^{-1}$ and $+10$ km s$^{-1}$ extracted from the CfA 1.2~m CO Survey Archive  \citep{dame01}. Contours levels are at 3, 4, and 5 K~km s$^{-1}$. The systemic velocity of the CO emission is about $-$7~km s$^{-1}$.}
      \label{intro}
\end{figure}

Sh2$-$236, surrounding the open cluster NGC~1893, is a semi-shell like \ion{H}{ii} region of about 55$\arcmin$ in size centered at RA=05:22:36.0; dec.=$+$33:22:00 (J2000) \citep{sharp59}. \citet{bli82}, based on CO (1$-$0) observations, found molecular gas related to Sh2$-$236 at the systemic velocity V$_{\rm LSR}$=$(-7.2 \pm 0.5)$~km~s$^{-1}$. To roughly
appreciate the distribution of the molecular gas at the CO (1--0) line, see the yellow contours in Fig.\,\ref{intro}.
 NGC~1893 is a well-studied open cluster which consists of at least five O-type stars \citep[][]{ave84,mar02}. The systemic velocity of this cluster was estimated to be $(-3.9 \pm 0.4)$~km~s$^{-1}$ \citep{lim18}. 

The most striking features in the region are two CGs located towards the northeast of Sh2$-$236: Sim~129 and Sim~130 (hereafter S129 and S130), whose heads point towards the location of the stars HD~242935, BD+33 1025, and HD~242908 (see Fig.\,\ref{fig2}). \citet{lim18} suggested that these stars are the main ionization sources of S129 and S130. Table \ref{stars} shows their spectral type and radial velocity (RV) at V$_{\rm LSR}$. Regarding the ionized gas associated with these structures, which can be traced by the radio continuum emission, 
it is worth noting that the NVSS sources 052307+332832 and 052255+333156 \citep{con98} coincide with S129 and S130, respectively.

\begin{table}
\caption{Ionizing stars of S129 and S130. Columns 2 and 3 indicate the spectral type and the radial velocity, respectively. The spectral types of HD 242935 was obtained by \citet{neg07} and of  BD +33 1025 and HD~242908 by \citet{sota14}. The radial velocities were obtained from Lim et al. (2018).}
\centering
\begin{spacing}{1.5}
\begin{tabular}{ccc}
\hline 
{\bf Name} & {\bf Spectral type}  &  {\bf V$_{\rm LSR}$ km s$^{-1}$}\\
\hline
HD~242935 & O7.5V & -2.2 $\pm$ 3.0\\
BD +33 1025 & O7.5V & -1.9 $\pm$ 8.6\\
HD~242908 & O4.5V & -3.8 $\pm$ 0.8\\
\hline
\label{stars}
\end{tabular}
\end{spacing}
\end{table}

\citet[][and references  therein]{lim18}  show evidence of a feedback-driven star formation related to NGC~1893. Despite a confusion with the names of the CGs (the names of Sim 129 and 130 are inverted in their work), they found several young stars, likely triggered by the action of the main cluster members. These young stars, spatially distributed in a region located between the center of the cluster and the CGs, show an age gradient, strongly suggesting that sequential star formation occurs in Sh2$-$236 \citep[see also][]{mah07}. According to \citet{lim18}, the RV of these young stars are in agreement with the RV of S129. 
In summary, \citet{lim18} found two peaks in the RVs distribution associated with the stellar population in the region, one at the V$_{\rm LSR} \sim -4 $ km~s$^{-1}$, and other one at about $+$2~km~s$^{-1}$, corresponding to the systemic velocity of the main star cluster and to the RV of the young stars likely related to S129, respectively. Finally, the authors, based on optical and near-IR spectroscopic  observations of the ionized and neutral gas, pointed out that S129 and S130 lie on different parts of an expanding bubble moving away from the cluster center. 

Summarizing, several works have studied the region of S129 and S130 in Sh2$-$236, but with the exception of \citet{lim18}, they focused on the stellar content. Therefore, for a complete understanding of the processes that are occurring in the region, a study of the molecular gas associated with the globules is required. In this work, based on new molecular observations, we present a kinematic study of the molecular gas associated with  S129 and S130 and their environments. Additionally, public multi-wavelength data were used to complement the molecular study.


\section{Molecular observations and data reduction}

The observation of the the molecular lines was carried out on October 3, 4, and 5, 2016 with the 10~m Atacama Submillimeter Telescope Experiment \citep[ASTE,][]{eza04}. We used the  DASH~345~GHz band receiver, which is a cartridge-type dual$-$polarization side$-$band separating (2SB) mixer receiver for 350~GHz band remotely tunable in the LO frequency range of 330-366~GHz. We simultaneously observed $^{12}$CO J=3$-$2 at 345.79~GHz and HCO$^+$J=4$-$3 at 356.734~GHz, mapping two regions of about $1\arcmin \times 1\arcmin$ centered at RA=05:23:06; dec.= $+$33:28:37 and RA=05:22:58; dec.= $+$33:31:35 (J2000) ( blue squares in Fig. \ref{fig2}), which mainly include the CGs S129 and S130, respectively and others interesting nearby structures. The mapping grid spacing was $20\arcsec$, and the integration time was 60 seconds per pointing. We also performed two single pointings centered at RA=05:23:08; dec.= $+$33:28:31 and RA=05:22:55; dec.= $+$33:31:40 (J2000) (green crosses in Fig. \ref{fig2}) of C$_2$H N=4$-$3 at 349.34~GHz and HNC J=4$-$3 at 362.63~GHz simultaneously, and HCN J=4$-$3 at 354.50~GHz  with an integration time of about 30 minutes per pointing. 

The observations were performed in position$-$switching mode. We used the XF digital spectrometer MAC with a bandwidth and spectral resolution set to 128~MHz and 125~kHz, respectively. The velocity resolution is 0.11~km~s$^{-1}$ and the beam size (FWHM) is $22\arcsec$ at 350~GHz. The system temperature varied from about T$_{sys}$=150 to 250~K during the observations. The main beam efficiency was $\eta_{mb}\sim 0.65$.
The data were reduced with NEWSTAR\footnote{Reduction software based on AIPS developed at NRAO, extended to treat single dish data with a graphical user interface (GUI)} and the spectra processed using the XSpec software package\footnote{XSpec is a spectral line reduction package for astronomy which has been developed by Per Bergman at Onsala Space Observatory}. The baseline fitting was done using second order polynomials for all lines. The HCO$^+$, HCN, HNC, and C$_2$H J=4$-$3 spectra were Hanning-smoothed to improve the signal-to-noise ratio.  The resulting spectral resolution of the observations was about 0.4~km~s$^{-1}$. The rms noise level  was about 100~mK for $^{12}$CO J=3$-$2 and HCO$^+$ J=4$-$3 transitions, and about 30~mK for the rest of the molecules.

\begin{table} 
\caption{Molecules observed with ASTE. Columns 2 and 3 indicate the detection or not of the molecule (transition) towards S129 and S130, respectively.} 
\centering
\begin{spacing}{1.5}
\begin{tabular}{ccc}
\hline 
{\bf Molecule (Transition)} & {\bf S129}  &  {\bf S130}\\
\hline
$^{12}$CO (J=3$-$2) & yes & yes\\
HCO$^+$ (J=4$-$3) & no & yes\\ 
HCN (J=4$-$3) & no & yes \\ 
HNC (J=4$-$3) & no & yes \\
C$_2$H (N=4$-$3, J=9/2--7/2) & yes & yes\\
\hline
\label{molecules}
\end{tabular}
\end{spacing}
\end{table}
   
\section{Results}

\subsection{Description of the infrared and optical emission}
\label{op}

\begin{figure}
\centering
 \includegraphics[width=9.2cm]{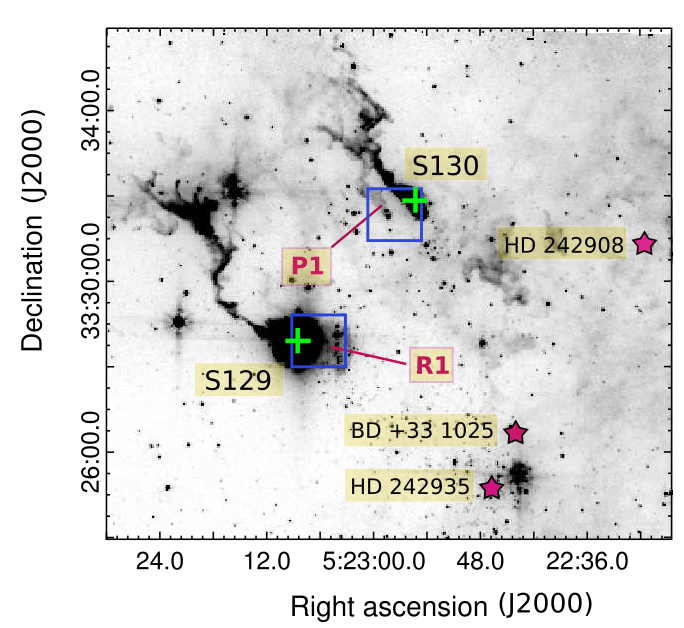}
 \caption{{\it Spitzer}-IRAC emission at 5.8~$\mu$m of  S129 and S130. The greyscale goes from 3 to 9 MJy/sr. P1 and R1 indicate the position of a pillar-like feature and a region of diffuse gas, respectively. The blue squares indicate the regions mapped in $^{12}$CO J=3$-$2 and HCO$^+$ J=4$-$3 lines. The green crosses represent the location of the single pointings  
 of the HCO$^+$, HCN, HNC J=4$-$3, and C$_2$H N=4$-$3 J=9/2--7/2 lines.}
      \label{fig2}
\end{figure}

Figure\,\ref{fig2} shows a zoom-up view of the region containing  S129 and S130 as seen at 5.8~$\mu$m  extracted from the {\it Spitzer}/GLIMPSE Survey \citep[][]{ben03, chu09}. The green crosses indicate the positions of the single pointings observations of the HCN and HNC J=4$-$3 and C$_2$H N=4$-$3 J=9/2--7/2 line. The blue squares indicate the regions mapped in the $^{12}$CO J=3$-$2 and HCO$^+$ J=4$-$3 transitions. These regions include S129 and S130, but also two striking structures, which are presented for the first time in this work. The first one, is a small pillar-like feature that appears in projection close to S130 (P1 in Fig.\,\ref{fig2}). The second one, can be appreciated as  the diffuse emission at 5.8~$\mu$m (R1 in Fig. \ref{fig2})  that is located, in projection, in front of the head of S129. 

\begin{figure*}
\centering
 \includegraphics[width=18cm]{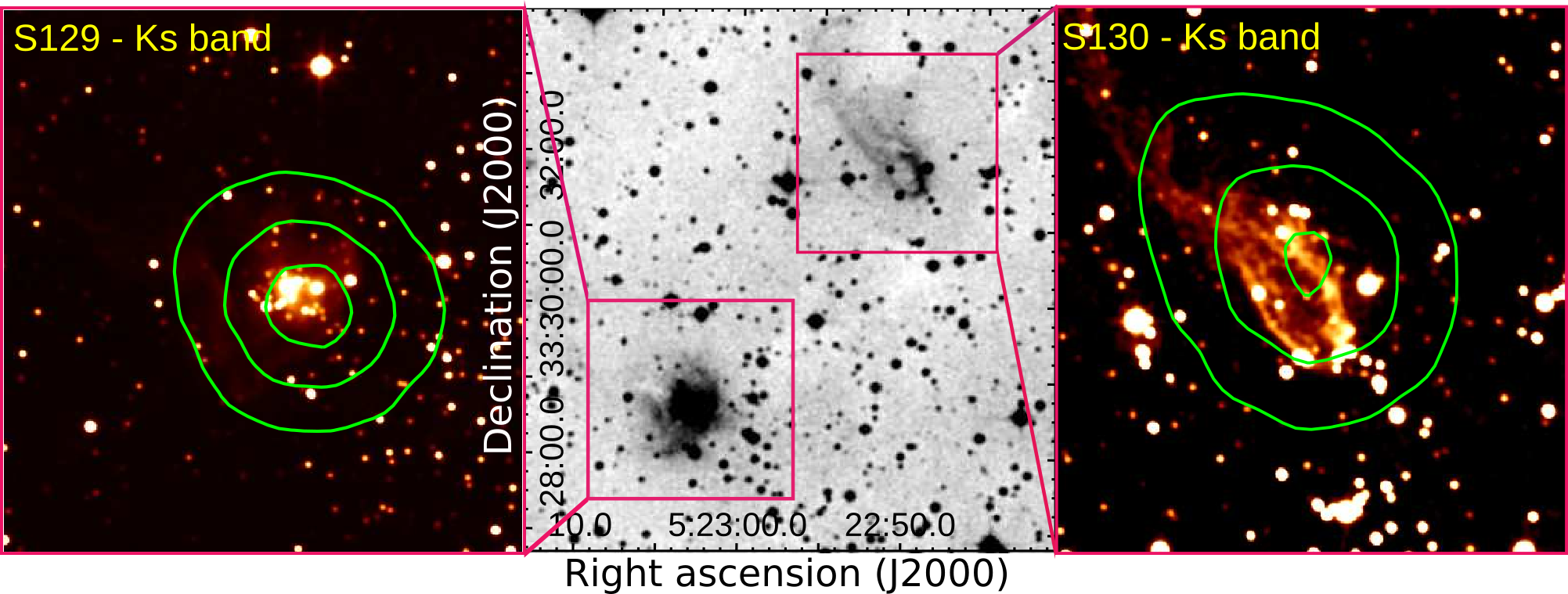}
 \caption{Central-panel: Optical emission extracted from the Digital Sky Survey 2 (DSS2)  towards  S129 and S130. Left-panel: Zoom-up view of S129 at the UKIDSS $K_s$ band. Right-panel: Zoom-up view of S130 at the UKIDSS $K_s$ band. The green contours represent the radio continuum emission at 1.4~GHz extracted from the NRAO VLA Sky Survey \citep[NVSS;][]{con98}. Levels are at 5, 10, and 15~mJy~beam$^{-1}$. The radio continuum shows the NVSS sources 052307+332832 and 052255+333156 related to S129 and S130, respectively.}
      \label{fig3}
\end{figure*}

Figure\,\ref{fig3}-central panel shows the optical emission towards S129 and S130 extracted from the Digital Sky Survey\footnote{http://archive.eso.org/dss/dss}. Both CGs exhibit conspicuous bright rims (BRs) facing to the location of the ionizing stars. The orientation of the BR related to S130 suggests that the star HD~242908 is the main responsible for ionizing the globule. In particular, towards S129 it can be noticed a striking  nebulosity located, in projection,  at the interior of the globule. This nebulosity could be associated with the young stars cluster found by \citet{mah07}. Figure\,\ref{fig3}-left and -right panels show the K$_s$ band emission extracted from UKIDSS survey\footnote{http://www.ukidss.org/surveys/surveys.html} towards S129 and S130, respectively. The green contours represent the radio continuum emission at 1.4~GHz obtained from the NVSS\footnote{https://www.cv.nrao.edu/nvss/} which traces the ionized gas associated with the CGs. Despite the relatively low angular resolution of the survey ($\sim45\arcsec$), its positional uncertainty of about 1 arcsec allows us to conclude that the peak of the radio continuum emission related to S129 coincides with the position of the over-density of stars likely embedded in the globule, suggesting that some of these young stars have begun to ionize the inner gas of S129. However, we can not discard a contribution of the ionized boundary layer to the radio continuum emission towards this globule. On the other hand, the radio continuum emission associated with S130 seems to peak at the brightest border of the globule, which suggests that this emission mostly arises from its ionized boundary layer. The elongated morphology of the radio continuum emission towards S130 is consistent with a globular cometary shape, reinforcing the scenario in which the ionized gas is related to the action of external sources.

\subsection{The molecular gas}
\subsubsection{$^{12}$CO J=3$-$2 spectra}
\label{co}

\begin{figure}
\centering
 \includegraphics[width=9cm]{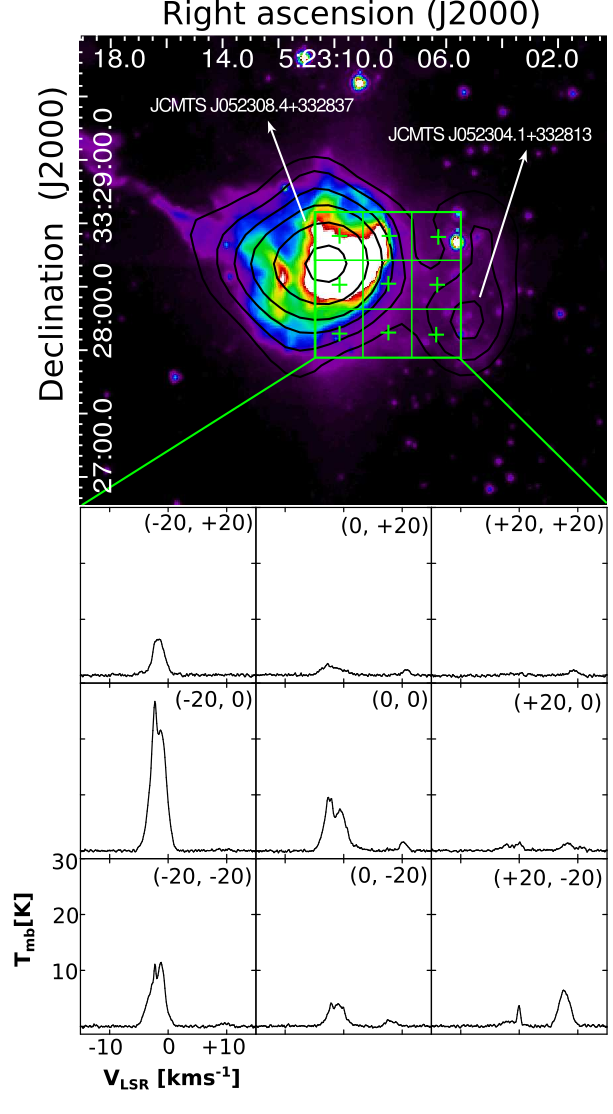}
 \caption{$^{12}$CO J=3$-$2 spectra towards the region of S129. The region where the spectra were taken is shown superimposed on the {\it Spitzer}-IRAC 5.8~$\mu$m emission.  The colour scale goes from 10~MJy~sr$^{-1}$ (violet) to 60~MJy~sr$^{-1}$ (white). The black contours represent the sub-millimeter continuum emission at 850~$\mu$m as extracted from the SCUBA survey. The levels are at 40, 70, 100, 120, 180, and 250 mJy~beam$^{-1}$.}
      \label{fig4}
\end{figure}

Figure\,\ref{fig4} shows  the $^{12}$CO J=3$-$2 spectra obtained with ASTE towards S129 (bottom panel) and the region where these spectra were taken (green squares and crosses)  superimposed on the {\it Spitzer}-IRAC 5.8~$\mu$m emission (top panel). The black contours represent the submillimeter continuum emission at 850~$\mu$m from the Submillimeter Common-User Bolometer Array \citep[SCUBA;][]{dif08} extracted from the James Clerk Maxwell Telescope (JCMT) data archive\footnote{www.jach.hawaii.edu/JCMT/archive/}. The SCUBA sources JCMTS J052308.4$+$332837 and JCMTS J052304.1$+$332813, catalogued by \citet{dif08}, positionally coincides with the head of S129 and with the region of diffuse gas identified as R1 in Fig.\,\ref{fig2}, respectively. The mapped region of about $1\arcmin \times 1\arcmin$  partially covers the location of both SCUBA sources, which seem to be connected in projection. Towards the ($-$20, 0) offset  the spectrum shows two velocity components centered at about $-$2.6~km s$^{-1}$ and at $-$0.9~km s$^{-1}$. These components also appear towards other offsets, but weaker. As we move from the head of S129 towards R1, a third velocity component appears centered at about $+$10.2~km s$^{-1}$. This component is more intense at the ($+$20, $-$20) offset, which suggests that it corresponds to molecular gas related to R1. 
Towards the (0, 0) offset  the spectrum shows three velocity components centered at about $-$2.5~km s$^{-1}$, $-$0.7~km s$^{-1}$ (the same that at the ($-$20, 0) offset), and $+$10.2~km s$^{-1}$ (likely related to R1 feature).

Table\,\ref{12co} presents the parameters derived from the Gaussian fittings to the spectra at the ($-$20, 0) and (0, 0) offsets, i.e. the positions that coincide with S129 head.  

\begin{figure}
\centering
 \includegraphics[width=9cm]{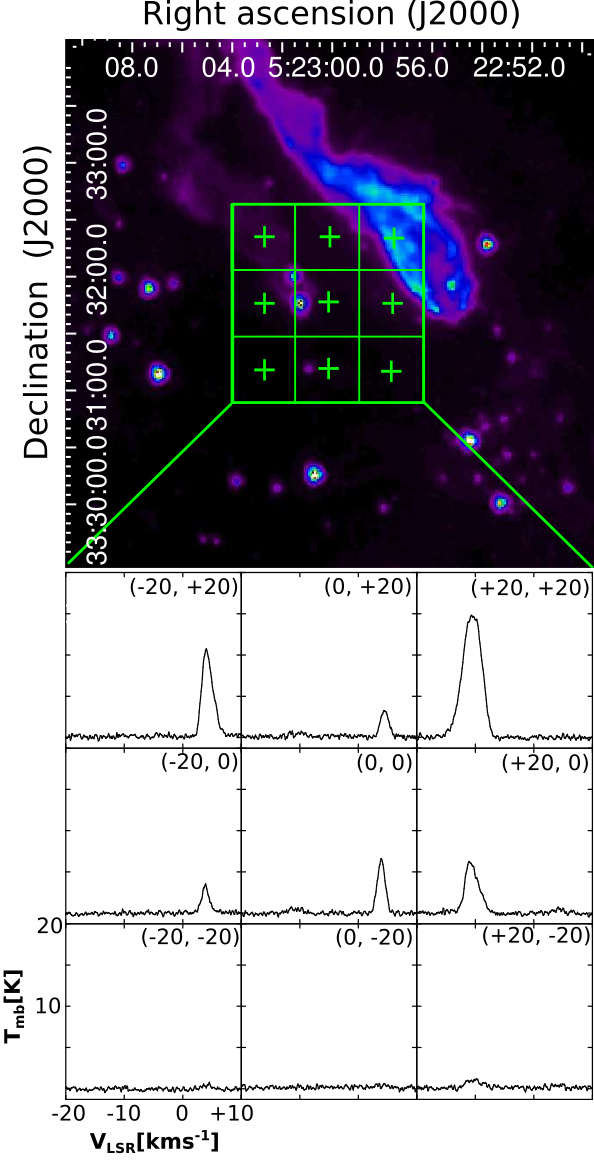}
 \caption{$^{12}$CO J=3$-$2 spectra towards the region of S130. The region where the spectra were taken is shown superimposed on the {\it Spitzer}-IRAC 5.8~$\mu$m emission.  The colour scale goes from 10~MJy~sr$^{-1}$ (violet) to 50~MJy~sr$^{-1}$ (green).}
      \label{fig5}
\end{figure}

Figure\,\ref{fig5} shows  the $^{12}$CO J=3$-$2 spectra obtained with ASTE towards  S130 (bottom panel) and the region where these spectra were taken (green squares and crosses)  superimposed on the {\it Spitzer}-IRAC 5.8~$\mu$m emission (top panel). The mapped region of about $1\arcmin \times 1\arcmin$  partially covers the head of S130 and the small 
pillar-like feature P1. The spectra at the ($+$20, $+$20) and ($+$20, 0) offsets, which were taken  in positional coincidence with S130,  show a velocity component centered at about $-$11~km~s$^{-1}$. This component corresponds to molecular gas associated with the CG. By the other hand, the spectrum at the ($-$20, $+$20) offset shows a velocity component centered at about $+$4~km~s$^{-1}$. This component is also present at the (0, $+$20), ($-$20, 0), and (0, 0) offsets, but weaker. Given the positions of these spectra, this velocity component would correspond to molecular gas associated with the small pillar-like feature P1.

\begin{table} 
\caption{Parameters derived from Gaussian fittings to the $^{12}$CO spectra of S129 at ($-$20, 0) and (0, 0) offsets (see Fig.\,\ref{infall}).} 
\centering
\begin{spacing}{1.5}
\begin{tabular}{ccc}
\hline 
 & {\bf S129} & \\
\hline
T$_{\rm mb}$ [K] & V$_{\rm LSR}$ [km~s$^{-1}$] & $\Delta$v [km~s$^{-1}$]\\
\hline 
\multicolumn{3}{c}{\bf  ($-$20, 0) offset}\\
\hline
 20.2$\pm$2.4  &  -0.9$\pm$0.2  &  2.2$\pm$0.2 \\ 
 25.3$\pm$2.2  &  -2.6$\pm$0.4  &  2.1$\pm$0.2 \\
\hline 
\multicolumn{3}{c}{\bf (0, 0) offset}\\
\hline
1.5 $\pm$0.2  & 10.2$\pm$1.8  &  1.4$\pm$0.2 \\ 
0.8 $\pm$0.1  &  1.7$\pm$0.2  &  2.7$\pm$0.5(*)\\
6.8 $\pm$0.8  & -0.7$\pm$0.1  &  1.8$\pm$0.2 \\
9.2 $\pm$1.2  & -2.5$\pm$0.3  &  1.9$\pm$0.3 \\
\hline 
\multicolumn{3}{l}{(*) Red wing candidate}\\
\label{12co}
\end{tabular}
\end{spacing}
\end{table}

\subsubsection{High-density molecular gas tracers}
\label{hco+}

\begin{figure}
\centering
 \includegraphics[width=9cm]{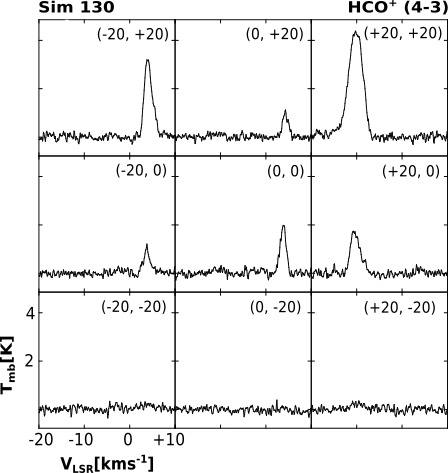}
 \caption{HCO$^+$ J=4$-$3 spectra towards the region of S130. The region where the spectra were taken is the same as shown in Fig.\,\ref{fig5}.}
      \label{fig5b}
\end{figure}

The HCO$^+$ molecule was only detected towards S130. Figure\,\ref{fig5b} shows the HCO$^+$ J=4$-$3 spectra, which were taken towards the same region that the $^{12}$CO spectra (see Fig.\,\ref{fig5}). The same two velocity components that were observed in the $^{12}$CO spectra at $-$11~km~s$^{-1}$ (S130) and at $+$4~km~s$^{-1}$ (P1) can be appreciated.

\begin{table} 
\caption{Parameters derived from the Gaussian fitting to the HCN, HNC J=4--3, and C$_2$H N=4--3 J=9/2--7/2 spectra shown in Figs.\,\ref{fig8} and\,\ref{c2h}.} 
\centering
\small
\begin{spacing}{1.5}
\begin{tabular}{cccc}
\hline 
Molecule (4--3)  &  T$_{\rm mb}$ [K] & V$_{\rm LSR}$ [km~s$^{-1}$] & $\Delta$v [km~s$^{-1}$]\\
\hline 
\multicolumn{4}{c}{\bf S129}\\
\hline
C$_{2}$H   & 0.08 $\pm$ 0.03 & -3.6 $\pm$ 0.2 & 2.9 $\pm$ 0.4\\ 
                   & 0.06 $\pm$ 0.03 & -0.8 $\pm$ 0.2 & 2.8 $\pm$ 0.4 \\
\hline 
\multicolumn{4}{c}{\bf S130}\\
\hline
C$_{2}$H & 0.20 $\pm$ 0.03 & -11.5 $\pm$ 0.3 & 1.8 $\pm$ 0.3\\
HCN   & 0.74 $\pm$ 0.05 & -10.8 $\pm$ 0.4 & 2.5 $\pm$ 0.2\\ 
HNC   & 0.25 $\pm$ 0.06 & -11.3 $\pm$ 0.3 & 1.9 $\pm$ 0.2\\
\hline
\label{HDT}
\end{tabular}
\end{spacing}
\end{table}

\begin{figure}
\centering
\includegraphics[width=8.5cm]{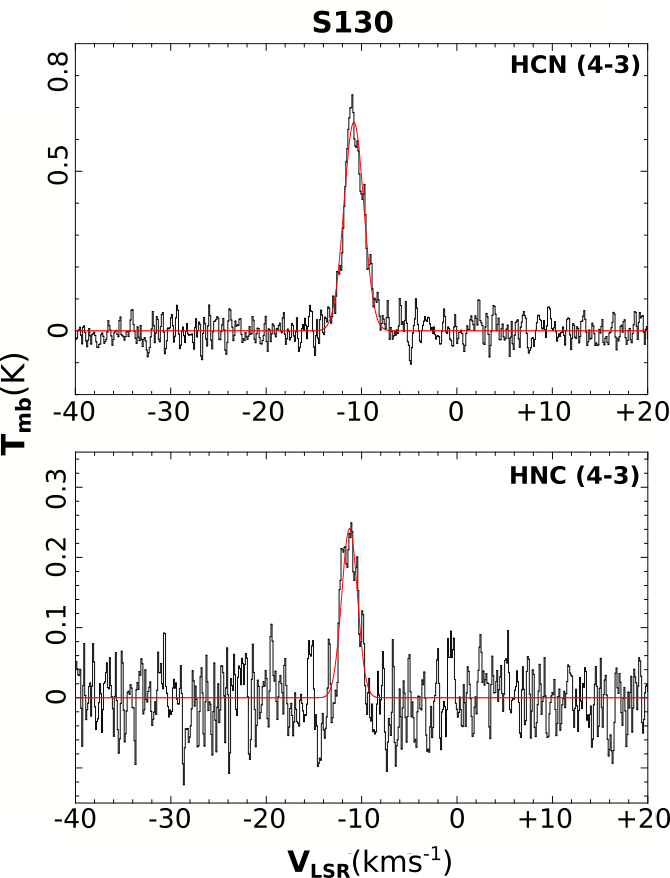}
 \caption{HCN J=4$-$3 (upper-panel) and HNC J=4$-$3 (lower-panel)  spectra towards  S130. The spectra were obtained towards the position of the green cross shown in Fig. \ref{fig2}. The red curves correspond to Gaussian fittings. The derived parameters are shown in Table \ref{HDT}. }
      \label{fig8}
\end{figure}

Figure\,\ref{fig8} shows the HCN J=4$-$3 (upper-panel) and HNC J=4$-$3 (lower-panel) spectra obtained towards the head of S130 (green cross at the S130 position  in Fig.\,\ref{fig2}). Both lines show the same velocity component centered at about $-$11~km s$^{-1}$ as seen in the $^{12}$CO line towards the ($+$20, $+$20) and ($+$20, 0) offsets (see Fig.\,\ref{fig5}). The parameters obtained from Gaussian fittings are listed in Table\,\ref{HDT}. The  HCN/HNC integrated ratio is about 3. HCN and HNC were not detected in S129.

\begin{figure}
\centering
\includegraphics[width=8.5cm]{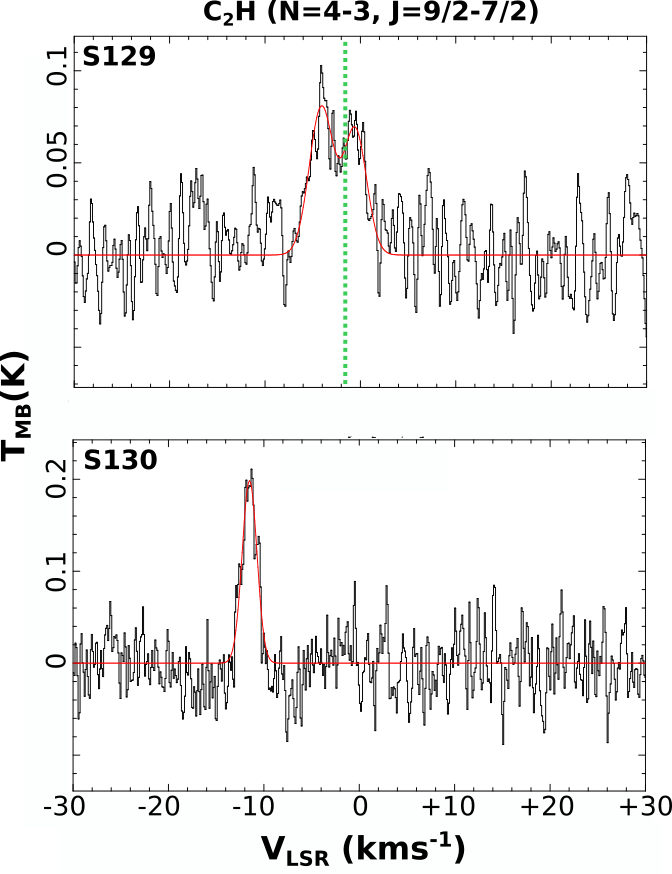}
  \caption{C$_2$H N=4$-$3 J=9/2--7/2 spectra towards S129 (upper-panel) and S130 (lower-panel). The spectra were obtained towards the positions of the green crosses shown in Fig. \ref{fig2}. The red curves correspond to Gaussian fittings. The derived parameters are shown in Table \ref{HDT}. The dashed vertical green line indicates the RV for the H$_2$ 1$-$0 S(1) line related to S129 found by \citet{lim18}.}
      \label{c2h}
\end{figure}

\subsubsection{C$_2$H molecule: a photo-dissociation region tracer}  

The C$_2$H was the only molecule, with the exception of $^{12}$CO, detected towards both CGs. Figure\,\ref{c2h} shows the C$_2$H N=4$-$3 J=9/2--7/2 spectra obtained towards S129 and S130 (green crosses in Fig.\,\ref{fig2}). The spectrum obtained towards S129 shows two velocity components centered at about $-$3.6~kms$^{-1}$ and $-$0.8~kms$^{-1}$. The shape of the profile is similar to that of the $^{12}$CO J=3$-$2 at the ($-$20, 0) offset towards S129, but the central velocity of the blue-shifted components differs in  1~km s$^{-1}$. The C$_2$H spectrum obtained towards S130 exhibits just one velocity component centered at about $-$11~km s$^{-1}$, which corresponds to the same component detected in the $^{12}$CO J=3$-$2 and HCO$^+$ J=4--3 at the ($+$20, $+$20) offset. The Gaussian fitted parameters are listed in Table\,\ref{HDT}.

\section{Discussion}

In this section we discuss the more relevant spectral features  presented above in order to: (1) unveil the nature of each individual object, and (2) 
disentangle the spatial distribution of the sources in the context of Sh2--236.

\subsection{Infall signatures and molecular outflows in S129}
\label{outflow}

An analysis of the molecular line profiles may help  us to identify kinematic  signatures associated with different dynamic processes that could be occurring in the interior of the molecular clouds, such as outflows, rotation, expansion and infall. The inwards motion of gas caused by gravity in regions of star formation, for instance, may lead to a self-absorbed emission line profile of an optically thick line. Simple early models \citep[see e.g.][and references therein]{mye96} predict a `blue asymmetry’ where the blue peak in the double-peak profile becomes brighter than the red one. Several  studies found observational evidence supporting this model \citep[e.g.][]{lee04, tsa08, sta10}.

\begin{figure}
\centering
 \includegraphics[width=9cm]{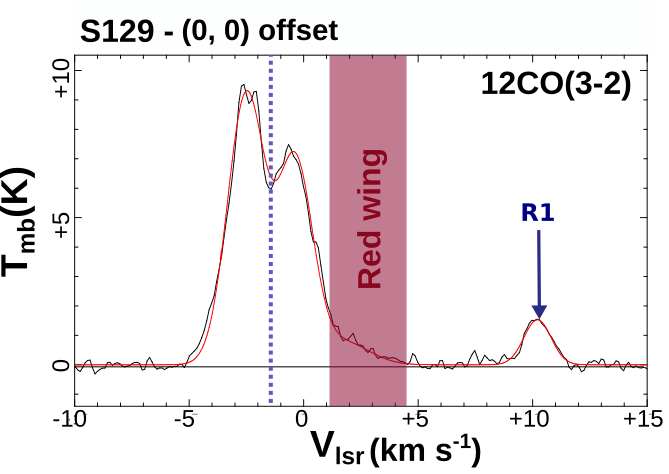}
 \caption{$^{12}$CO J=3$-$2 spectrum towards S129 at the (0, 0) offset. The red curve correspond to the Gaussian fitting. The dashed vertical blue line indicates the RV for the H$_2$ 1$-$0 S(1) line related to S129 found by \citet{lim18}.}
      \label{infall}
\end{figure}

Figure\,\ref{infall} shows the $^{12}$CO J=3$-$2 spectra towards S129 at the ($-$20, 0) and (0, 0) offsets. The blue component appears more intense than the red one, which may imply infall motions of the gas. 
The C$_2$H J=4$-$3 spectrum towards the head of S129 shows the same `blue-skewed' profile (see Fig.\ref{c2h}). Thus, it is likely that the `dip' observed at $-$1.5~km s$^{-1}$ and at $-$2~km s$^{-1}$ in the $^{12}$CO and C$_2$H profiles, respectively, is a self-absorption feature. Moreover, \citet{lim18} detected hydrogen molecular gas towards S129 with a velocity component at about $-$1.5~kms$^{-1}$, which supports  the `dip' interpretation of the profiles and reinforces the scenario of infall gas motion in the CG. 

On the other hand, the $^{12}$CO J=3$-$2 spectrum  presented in Fig.\,\ref{infall} shows evidence of turbulent gas. It can be noticed a red-wing at the (0, 0) offset, likely related to the molecular outflow activity generated by the young stars that are forming inside S129.

\subsection{Analyzing the high-density gas in the globules}

We detected HCO$^+$ J=4$-$3 only towards the region of S130. The results showed in Section \ref{op} suggest that, despite the low angular resolution of the NVSS data, the ionized gas related to S129 seems to have two contributions: one from the ionized boundary layer, and other one from the activity of young stars embedded in the globule, while most of the ionized gas associated with S130 seems to arise from the illuminated border of the globule. 
\citet{goi09} found a decreasing of the HCO$^+$ abundance from the shielded core to the UV irradiated cloud edge in the Horsehead nebula. The authors attributed this diminishing of the HCO$^+$ abundance to the electronic recombination, which becomes more important as the number of free electrons increases. The angular resolution of our molecular observations does not allow us to discern whether the HCO$^+$ emission arises from the interior of the globule or from the irradiated border of the cloud. However, it is likely that the HCO$^+$ emission detected towards S130 comes from the neutral gas of its interior, while the non-detection of HCO$^+$ towards S129 could be explained by the presence of ionized gas at both, interior and the edge of the globule. This points out to different star forming evolutionary stages between both globules \citep[][]{san12, hoq13}.

The HCN and HNC molecules are also high-density gas tracers.  As in the case of HCO$^+$, we only detected HCN and HNC J=4$-$3 lines towards S130. \citet{gran14} suggested that the HCN/HNC abundance ratio can be used to trace the evolutionary stages of star forming regions. \citet{jin15} found a statistically increasing tendency of the abundance ratio with the evolution of the objects, with values that go from about 1 in the starless clumps to about 9 in the UC\ion{H}{ii} regions.  According to the authors, this may be due to a temperature increase that favors the reaction: HNC + H $\xrightarrow{}$ HCN + H. In addition, \citet{che16} and \citet{aguado17} showed that in the interstellar regions exposed to intense UV radiation fields, HCN should be more abundant than HNC. They showed that the destruction of both isomers is dominated by the photodissociation and the HNC is destroyed faster than the HCN. Taking into account these results and assuming similar excitation conditions for both isomers towards the head of S130, which is a quite likely assumption, the integrated ratio of about 3 derived in Section\,\ref{hco+} is in agreement with a scenario of incipient star formation where the young stars are not ionizing the interior of the globule.
On the other hand, the non-detection of both isomers towards S129 could be due to a low abundance (emission undetectable for the ASTE sensitivity), or to that both isomers were destroyed by the radiation field from the likely cluster of young stars within this globule as seen in the $K_{s}$ band (Fig.\,\ref{fig3} left).

Finally, the C$_2$H N=4$-$3 J=9/2--7/2 line was detected towards S129 and S130. It is well known that the C$_2$H molecule is a photo-dissociation regions (PDRs) tracers \citep[][]{fue93, jan95, fue96, nagy15}. The PDR is the transition layer between the ionized gas directly irradiated by strong UV fields (e.g. from massive OB stars) and the cold neutral gas shielded from radiation \citep[e.g.][]{tie85}. \citet{tie13} present one of the most complete studies of the chemistry of this molecule involved in the PDRs. 

The C$_2$H N=4$-$3 J=9/2--7/2 towards S129 exhibits a similar blue-skewed spectrum as the $^{12}$CO spectra towards the ($-$20, 0) and (0, 0) offsets, supporting the interpretation of infall motions of the gas as was previously discussed. We suggest that while the  HCO$^+$, HCN, and HNC molecules exhibit the differences between the internal conditions of the gas of S129 and S130, the ubiquitous presence of the C$_2$H molecule towards both globules shows that the nearby O-B stars are irradiating the external layers of S129 and S130. 

\begin{table}
\caption{Radial velocities of the gas associated with the four molecular structures characterized in this work. (*) Radial velocity of the 'dip' feature in the $^{12}$CO J=3$-$2 and C$_2$H profiles.}
\centering
\begin{spacing}{1.5}
\begin{tabular}{cc}
\hline
{\bf Structure name} & {\bf V$_{\rm LSR}$(kms$^{-1}$)}\\
\hline
S129 & $-$1.5 (*) \\
S130 & $-$11 \\
R1 & $+$10 \\
P1 & $+$4 \\
\hline
\label{vel}
\end{tabular}
\end{spacing}
\end{table}

\subsection{Schematic view of the molecular features in the context of Sh2-236}

Based on spectroscopic optical observations, \citet{lim18} found a bimodal distribution in RV of the stars in the region.  The main peak is at $-$3.9~kms$^{-1}$ and corresponds to the systemic velocity of NGC~1893. The secondary peak corresponds to a second subgroup of young stars at $+$2.6~km s$^{-1}$. The authors also found a bimodal behaviour in the RV of the ionized gas. From the forbidden line \ion{N}{ii} ($\lambda$6584), they observed two velocity components at $-$13~km s$^{-1}$ and $+$4.2~km s$^{-1}$, which correspond to the near and far side, respectively, of a  bubble with an expansion velocity of about 8.6~km s$^{-1}$. Finally, based on spectroscopic observations of the near-infrared H$_2$ 1$-$0 S(1) line, they found RVs of about $-$1.5~km s$^{-1}$ and $-$11~km s$^{-1}$ for the hot molecular gas related to S129 and S130, respectively.

Based on our observations, we found similar RV components of the cold molecular gas associated with S129 and S130 (see Table \ref{vel}), supporting that S129 and S130 should be located at the far and the near side of the shell, respectively.
Additionally, the molecular gas related to the small pillar-like feature P1 has a RV of $+$4~kms$^{-1}$, which is exactly the same RV of the far side of the shell of the 
expanding bubble.
 
Finally, the RV of the molecular gas related  to R1 (about $+$10~km~s$^{-1}$) differs in more than 11~km~s$^{-1}$ with the RV of S129, which suggests that while S129 is located at the far side of the expanding shell, R1 would be placed  well beyond. 


\section{Summary}

In the literature there are several studies about the cometary globules S129 and S130 in Sh2$-$236, but with the exception of \citet{lim18}, they focused on the stellar content. Therefore, for a complete understanding of the processes that are occurring in the region, a study of the molecular gas associated with globules was needed, and thus 
we  present  a  kinematic  study  of  the  molecular  gas  associated with S129 and S130 and their environments. The main results are summarized as follows.

We found kinematic signatures of infalling gas in  the $^{12}$CO J=3$-$2 and C$_2$H N=4$-$3 J=9/2--7/2 spectra towards S129. 
We detected HCO$^+$, HCN, and HNC J=4$-$3 only towards S130. The integrated ratio of HCN/HNC of about 3 is in agreement with a scenario of incipient star formation where the young stars are not ionizing the interior of the globule. The non-detection of these molecules towards S129 could be due to the radiation field arising from the star formation activity inside this globule.  The ubiquitous presence of the C$_2$H molecule towards S129 and S130 evidences the action of the nearby O-B stars irradiating the external layer of both globules. 

Based on the mid-infrared 5.8~$\mu$m emission, we identified two new structures: (1) a region of diffuse emission (R1) located in front of the head of S129  and (2) a pillar-like feature (P1) placed besides S130. 
Based on the $^{12}$CO J=3$-$2 line, we found molecular gas associated with S129, S130, R1 and P1 at radial velocities of $-$1.5~kms$^{-1}$, $-$11~kms$^{-1}$, $+$10~kms$^{-1}$, and $+$4~kms$^{-1}$, respectively. Therefore, while S129 and P1 are located at the far side of the shell, S130 is placed at the near side. 

Finally, the RV of the molecular gas related to R1 differs in more than 11~km~s$^{-1}$ with the RV of S129, which suggests that while S129 is located at the far side of the expanding shell, R1 would be placed  well beyond. 


\begin{acknowledgements}
We thank the anonymous referee for her/his fruitful comments.  The ASTE project is led by Nobeyama Radio Observatory (NRO),
a branch of National Astronomical Observatory of Japan (NAOJ), in collaboration with University of Chile, and Japanese institutes including University of Tokyo, Nagoya University, Osaka Prefecture University, Ibaraki University, Hokkaido University, and the Joetsu University of Education. M.O., and S.P., are members of the Carrera del investigador cient\'ifico of CONICET, Argentina. M.B.A. is a doctoral fellow of CONICET, Argentina. This work was partially supported by grants awarded by CONICET, ANPCYT and UBA (UBACyT) from Argentina.
M.R. wishes to acknowledge support from FONDECYT(CHILE) grant No
1140839. M.O. is grateful to Dr. Natsuko Izumi for the support received during the ASTE observations.
     
\end{acknowledgements}

%
%

 \bibliographystyle{aa} 
 \bibliography{ref} 

\end{document}